# STGIC: a graph and image convolution-based method for spatial transcriptomic clustering

Chen Zhang[1,¶,*], Junhui Gao[2,¶], Lingxin Kong[1], Guangshuo cao[3,&], Xiangyu Guo[4,&], Wei Liu[5]

[1] School of Life Sciences and Biotechnology, Shanghai Jiao Tong University, Shanghai, China

[2] Department of Computer Science and Engineering, Shanghai Jiao Tong University, Shanghai, China

[3] State Key Laboratory of Public Big Data, College of Computer Science and Technology, Guizhou University, Guiyang

[4] Smart-Health Initiative, King Abdullah University of Science and Technology, Jeddah, Saudi Arabia

[5] Marine Science and Technology College, Zhejiang Ocean University, Zhoushan, China

* Corresponding author

Email: zhang83890@sina.com

[¶]These authors contributed equally to this work.

[&]These authors contributed equally to this work.

# Abstract

Spatial transcriptomic (ST) clustering employs spatial and transcription information to group spots spatially coherent and transcriptionally similar together into the same spatial domain. Graph convolution network (GCN) and graph attention network (GAT), fed with spatial coordinates derived adjacency and transcription profile derived feature matrix are often used to solve the problem. Our proposed method STGIC (**s**patial **t**ranscriptomic clustering with **g**raph and **i**mage **c**onvolution) utilizes an adaptive graph convolution (AGC) to get high quality pseudo-labels and then resorts to dilated convolution framework (DCF) for virtual image converted from gene expression information and spatial coordinates of spots. The dilation rates and kernel sizes are set appropriately and updating of weight values in the kernels is made to be subject to the spatial distance from the position of corresponding elements to kernel centers so that feature extraction of each spot is better guided by spatial distance to neighbor spots. Self-supervision realized by Kullback–Leibler (KL) divergence, spatial continuity loss and cross entropy calculated among spots with high confidence pseudo-labels make up the training objective of DCF. STGIC attains state-of-the-art (SOTA) clustering performance on the benchmark dataset of 10x Visium human dorsolateral prefrontal cortex (DLPFC). Besides, it's capable of depicting fine structures of other tissues from other species as well as guiding the identification of marker genes. Also, STGIC is expandable to Stereo-seq data with high spatial resolution.

# Introduction

Single-cell transcriptomics (SC) technique has been fully developed to promote the research of a myriad of fields in life science via its advantage of high resolution to make clear the transcriptomic profile in every single cell [1]. SC clustering requires grouping cells of the same type together with only gene transcription information [2]. Methods originally for community detection have been used for SC clustering, such as Leiden and Louvain [3, 4]. ST technique has also been rapidly developed in recent years. Compared with SC, it takes account of one more aspect of the spatial information of the sequencing unit, also referred to as a spot [5]. Spatial resolution of most existing ST technique is generally not so high as SC and the improvement of spatial resolution is usually realized at the cost of drop in the number of different kinds of transcripts captured in one spot. At present, 10x Visium [6, 7] and Slide-seq [8, 9] have been recognized as the most popular technique for ST sequencing. Slide-seq boasts higher spatial resolution of 10 μm to attain nearly cellular level than 10x Visium with the spatial resolution of 55 μm. Despite the inferiority of 10x Visium in terms of spatial resolution, it does much better in the capturing efficiency of each spot for the mentioned paradox between spatial resolution and capturing efficiency. Stereo-seq [10] is another technique with even higher spatial resolution than Slide-seq to achieve sub-cellular level, and also the capturing efficiency is still open to improve. Both 10x Visium and Stereo-seq have their spots arranged into regular lattices on the chips, as is contrast to Slide-seq.

ST clustering aims to divide all spots from a sample slice into spatial domains according to both spatial vicinity and gene transcription similarity. Each spatial domain consists of continuously distributed spots, all of which present relative coherence in transcription profile. Based on division of spatial domains, many important downstream works can be done such as detecting spatially variable genes (SVGs) whose expression distribution have significant correlation with their spatial locations. SVGs are often markers for specific structural components of tissues or function as regulator for signal transduction pathways [11]. Therefore, identification of SVGs is tremendously helpful to unearth the underlying biological mechanisms behind tissue development [12], onset and progression of diseases [13].

So far, ST datasets with high quality of cluster or cell type labels have still been very scarce since the experimental labeling process is terribly time-consuming and costly. The 10x Visium human DLPFC dataset [14] composed of 12 samples and owning generally acceptable annotated labels about spatial domains has enabled the test of various ST clustering algorithms about clustering performance. Hence, the DLPFC dataset has been taken as benchmark to measure the precision of cluster assignment generated by related algorithm.

Leiden and Louvain considering only gene transcription information had also been used frequently for ST clustering until the naissance of the deep learning method SpaGCN [15] in late 2021 which adopts GCN [16] with an adjacency constructed according to spots' spatial distances to extract features from information on both gene transcription and histological image. The method can be used for clustering and detecting SVGs. Shortly after SpaGCN, STAGATE [17] was reported as a comprehensive ST toolkit, which carries out spots clustering via GAT and auto-encoder framework [18, 19] to reconstruct the gene expression information, the spatial coordinates is used to identify neighborhood for each spot, but different from SpaGCN, it doesn't take cues from histological image. DeepST [20] reported in late 2022 takes advantage of a pre-trained neural network to extract feature from histological image and integrate it with gene expression information to construct a feature matrix. Moreover, Graph neural network auto-encoder and denoising auto-encoder are simultaneously used to extract the latent feature which is further taken by Leiden to generate clustering labels. The foregoing graph learning methods are based on single graph, however, STMGCN [21] goes beyond the custom by constructing two graphs from one sample. Besides the adjacency calculated as SpaGCN does, another adjacency with binary value 0 and 1 is derived from cosine similarity calculated with spatial coordinates and only the weights of the elements corresponding to the top 20 neighbors in the adjacency are conferred to the value 1. Attention mechanisms are used to fuse the embeddings from the two graphs to obtain the final embeddings for clustering.

Besides graph-based deep learning method, an image convolution-based method abbreviated as TESLA [22] is also developed SpaGCN's author to annotate cell types *in situ* on histological images for tumor tissues. TESLA is fed with a 2-channel image with one channel derived from the gray degree converted from histological image and another from the transcription information of marker genes of all spots. The input is processed by several blocks containing convolution layers with kernel size of 3*3, the output of these blocks is used to generate pseudo-labels to supervise itself with cross-entropy loss and the spatial continuity is ensured by the constraint that longitudinally and transversely neighbored pixels should be similar in embedding space. The main configuration of TESLA refers to an unsupervised image segmentation method [23], rendering TESLA the first trial of image-based deep learning method in ST, though not for identifying spatial domains. Noticeably the input image is not simply histological images, rather it is virtualized from gene expression information. ScribbleDom [24] is another image-based method employing the convolution framework of Inception to extract feature from the virtual image and it depends on the clustering algorithm mclust to provide pseudo-labels for training.

In addition to deep learning methods, Bayesian statistic methods also find their way in ST analysis among which BayesSpace [25] are the first reported paradigm and has made itself a representative of the application of Bayesian statistics to ST clustering.

Herein, we introduce STGIC to deal with ST clustering problem specifically for techniques adopting regular lattices on chips, such as 10x Visium and Stereo-seq. STGIC consists of AGC [26] for pre-clustering and DCF for clustering. AGC is quite different from common graph convolution mainly in terms of two aspects, on one hand, it doesn't depend on any trainable parameters, on the other hand, it works adaptively to detect the appropriate order of neighbors to aggregate for different graphs and is thus more likely to avoid over-smoothing and under-smoothing spots' embeddings to achieve a good clustering performance. AGC functions to provide high quality of pseudo-labels for pre-training of DCF. Unlike the convolution framework adopted in TESLA, DCF adopts different combinations of convolution kernels for 10x Visium and Stereo-seq. Special steps have been taken to ensure that feature extraction of every spot-corresponding pixel pays attention to only the neighboring spot-corresponding pixels within a certain spatial distance and neglect all those corresponding to no spots or far away, besides, the extent to which neighbors are paid attention to by a spot is the same among those equally distant from the spot. Training of DCF doesn't either mechanically follow what is adopted in the foregoing unsupervised image segmentation algorithm, as is embodied not only by the AGC assisted pre-training, but also the transferring of a self-supervision loss originally designed for embeddings generated by GCN to that extracted from the feature image output by DCF, more directions to compute spatial continuity and more strict restraints on the eligibility of spots being involved in calculating cross entropy.

Tests of STGIC with the 10x Visium DLPFC benchmark and other datasets of different tissues, species and sequencing techniques indicate the following points: (*i*) STGIC attains SOTA clustering performance on the benchmark with high mean and median adjusted Rand index (ARI). (*ii*) STGIC can delineate the structure of human and mouse brains and human breast cancers in fine-grained scale. (*iii*) STGIC is competent for 10x Visum and Stereo-seq owning regular lattices.

# Results

## Pipeline of STGIC to analyze the benchmark of 10x Visium DLPFC

Implementation of STGIC is divided into stages, including preparation of input for AGC and DCF, pre-clustering with AGC, pre-training and training of DCF (Fig 1).

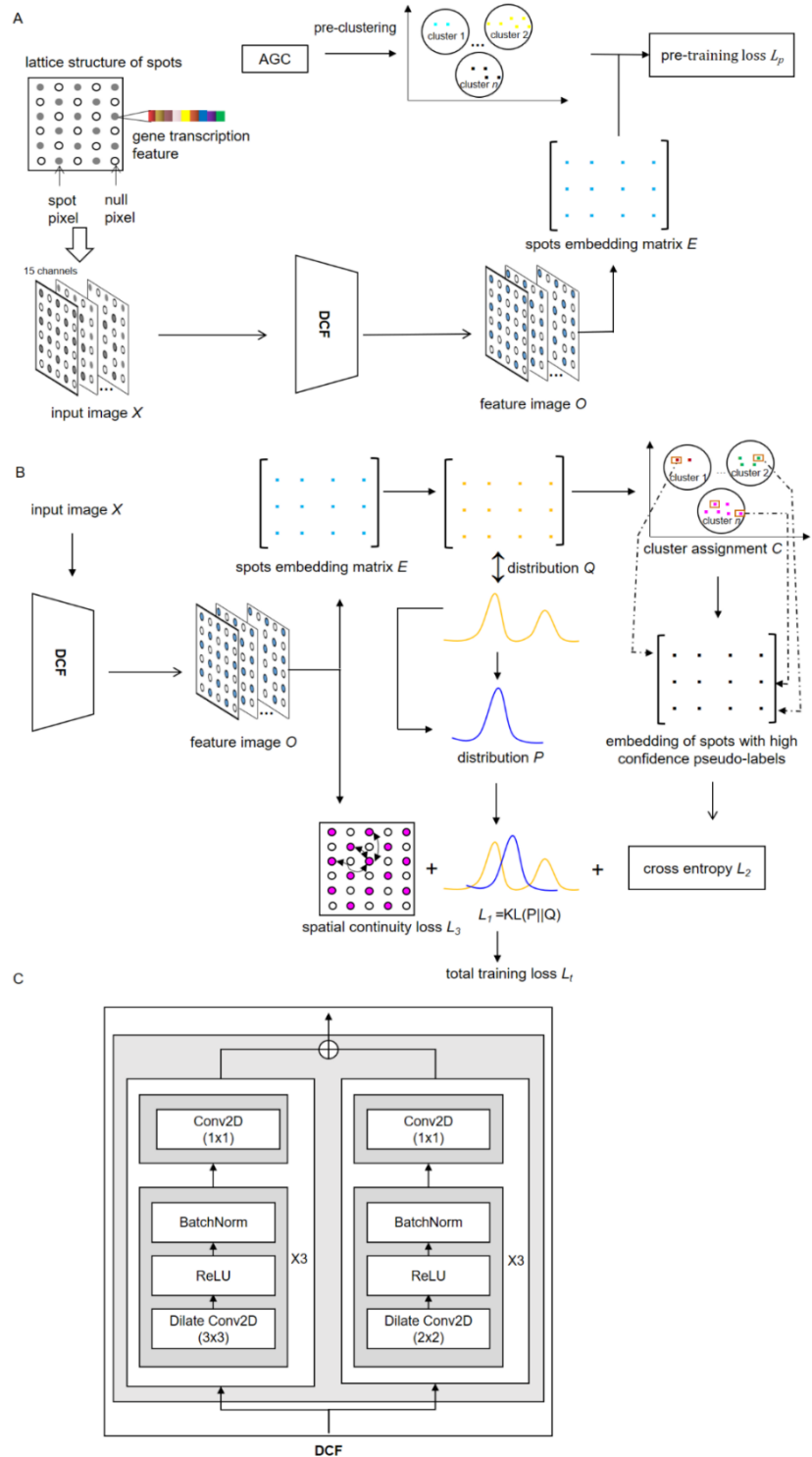

**Fig 1. Overview of STGIC.** A. Pre-training of DCF. A virtual image is converted from the lattice of 10x Visium and gene expression information to serve as the input of DCF. AGC works to generate a cluster assignment $C_0$ for supervising the pre-training of DCF. B. Training of DCF. Training of DCF starts with the trainable

parameters initialized by values identified during the pre-training stage and is carried out with objectives of minimizing the total loss consisting of three components. C. The detailed structure of DCF for 10x Visium. DCF is composed of two sub-frameworks depending on convolution kernels with dilation rate of 2 and size respectively of 2*2 and 3*3.

**A. Preprocessing of raw datasets.** Data of each sample are stored in a h5ad file from which a gene transcription matrix $M_0 \in R^{s_0 \times g_0}$ is extracted, $s_0$ represents number of spots and $g_0$ represents number of genes in unfiltered data. The matrix records unique molecular identifier (UMI) counts of genes detected in all spots. Besides, two kinds of coordinates are contained in 10x Visium data to identify positions of spot-corresponding pixels in the histological image and to index spots in the lattice, the two kinds of coordinates are termed pixel coordinates and lattice coordinates henceforth. Genes expressed in fewer than three spots are eliminated. The counts of gene expression of each spot is normalized by being divided with total UMI counts of all genes in the spot, multiplying 10,000 and taking natural logarithm [15], the resulting normalized gene expression matrix is denoted as $M \in R^{s \times g}$, where $s$ represents number of spots and $g$ represents number of genes in filtered data. The ST clustering task is to assign a cluster label to each spot and the cluster number $n$ is usually pre-specified.

**B. Preparation of inputs for the pre-clustering model AGC.** Pixel coordinates are used to compute the distances between each pair of spots whereby to construct a distance matrix $d \in R^{s \times s}$, based on which the adjacency $A \in R^{s \times s}$ is computed with Gaussian kernel and further a symmetrically normalized Laplacian matrix $L \in R^{s \times s}$ is computed. The top 50 principal components (PCs) are extracted via principal components analysis (PCA) of the matrix $M$ to get the feature matrix $F \in R^{s \times 50}$.

**C. Preparation of inputs for the clustering model DCF.** A virtual image $X \in R^{15 \times h \times w}$, is constructed with lattice coordinates and gene transcription information as the input for DCF, where the channel number is 15, $h$ and $w$ represent the height and width of the image. Each spot is arranged in the image as a pixel which is termed spot pixel, two closest neighbor spot pixels are separated by a pixel not corresponding to any spot and termed null pixel (Fig 2A). Surrounding spots pixels and null pixels are also pixels not corresponding to any spot but termed background pixels to differentiate from null pixels since there are indeed not any cells distributed in the area while there could have been cells in the positions of null pixels except for limitation of the experimental technique. The pixel values of spot pixels are derived by taking the top 15 PCs generated by PCA with the matrix $M$ and those of null and background pixels are also imputed.

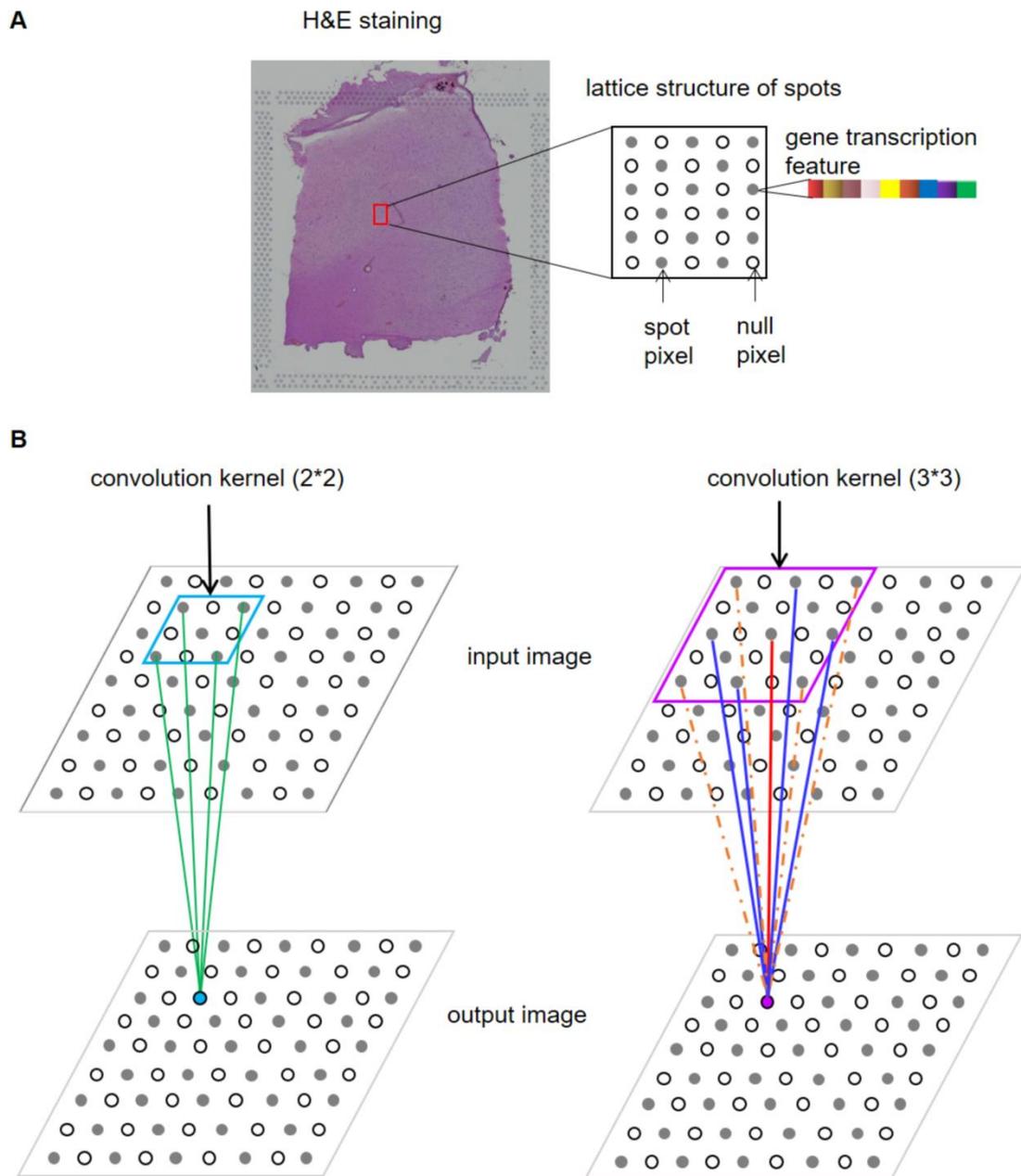

**Fig 2. Lattice structure of 10x Visium and the adaptation of convolution kernel to the structure.** A. Spots are arranged alternately on a chip to form a regular array. B. Two kinds of Convolution kernels are used in DCF. Both have dilation rate of 2, but one has kernel size of 2*2 and the other has that of 3*3. Convolution kernel shares the same weight at positions with equal distance to the center. Lines with the same color represent the same kernel weight when extracting feature from a receptive field. Orange dash lines represent the weight of zero to ignore these spots

during feature extraction.

**D. Pre-clustering with AGC.** AGC works in a manner different from common GCN, as it generates embeddings every iteration for each spot through aggregating the feature of neighbors with a pre-defined non-trainable kernel, then spectral clustering is performed with the embeddings and correspondingly an intra-cluster distance is calculated. Once the intra-cluster distance in current iteration is lower than that in last iteration, it terminates iteration and presents the cluster assignment in last iteration as its final clustering result in the form of a matrix $C_0 \in Z^{l \times s}$ (Fig 3).

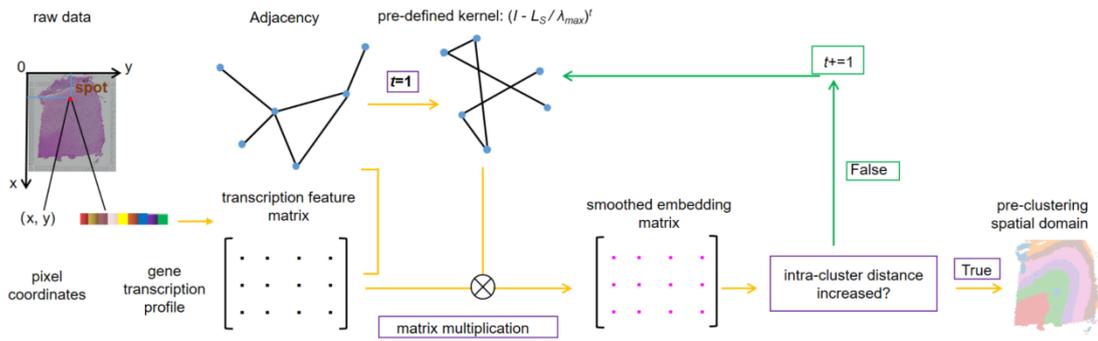

**Fig 3. Workflow of the pre-clustering method AGC in STGIC.** Pixel coordinates of spots is used to compute the Euclidean distance based on which adjacency is constructed and then the pre-defined kernel function can be calculated. The top 50 PCs of each spot serve as the beginning feature input and are aggregated with the guidance of the kernel function to obtain spots' embeddings for spectral clustering to generate a cluster assignment and the intra-cluster distance is subsequently calculated during each iteration. When the intra-cluster distance starts to increase, iteration terminates and the cluster assignment in the iteration presenting the minimum intra-cluster distance is adopted as the pre-clustering result.

**E. Feature Extraction with DCF for 10x Visium Data.** DCF relies mainly on two kinds of convolution kernel of size 2*2 and 3*3, both of which have the dilation rate set to be 2. Two sub-frameworks, both of which consist of 4 blocks are harnessed in parallel and the first 3 blocks of each sub-framework are repetition of a 3-layer unit for 3 times (Fig 1C). The 3-layer unit comprises sequentially the dilated convolution layer (kernel size 2*2 and 3*3 respectively in the two sub-frameworks), a batch normalization layer and a ReLU activation layer (Fig 1C). The last blocks of the two sub-frameworks contain only a convolution layer with kernel size 1*1. The outputs of the two sub-frameworks are summed up by pre-specified weights to generate a feature image $O \in R^{n \times h \times w}$, from which embedding matrix $E \in R^{s \times n}$ of all spots can be extracted according to the lattice coordinates (Fig 1C).

**F. Pre-training of DCF.** For one thing, a probability matrix is derived from $E$ via softmax function to represent the probability of each spot being assigned to each cluster (Fig 1A). For another, the pre-clustering labels serve herein as pseudo-labels. The cross-entropy loss $L_p$ can be computed with the probability matrix and the pseudo-labels (Fig 1A). The objective of the pre-training is to minimize the cross-entropy loss.

**G. Training of DCF.** Besides the initialization of all trainable parameters with the values ascertained during the pre-training stage, the centroid representation matrix $\mu \in R^{n \times n}$ is an extra trainable parameter declared in the stage, where the first $n$ refers to the cluster number and the second $n$ demand the representation vector for each centroid should also be $n$-dimensional, thus enabling the computation of the distance between the embedding of each spot and the representation of each cluster centroid to get the probability distribution matrix $Q \in R^{s \times n}$ (Fig 1B). Another probability distribution $P \in R^{s \times n}$ can be computed from $Q$ so that the minimization of KL divergence $L_1$ between the two distributions form a part of the training objective. $P$ is kept unchanged relatively to $Q$ by updating the value of $P$ matrix every 4 times of iterations. Besides, every spot is conferred to a label corresponding to the cluster with the highest probability according to the matrix of $Q$, whereby a cluster assignment $C \in Z^{1 \times s}$ for all spots can be generated to provide pseudo-labels in each iteration. Pseudo-labels of high confidence can be picked out if the corresponding probability is higher than $q_{cut}$ (here $q_{cut} = 0.5$). A cross entropy loss $L_2$ can be calculated among spots with high confidence pseudo-labels and the corresponding probability (Fig 1B). Besides, the spatial continuity loss $L_3$ is computed as the mean value of Euclidean distances between all pairs of closest neighboring spots in longitudinal, transverse and diagonal directions in the lattice based on spots' embeddings in the matrix $E$ (Fig 1B). The total loss $L_t$ equals the weighted sum of $L_1$, $L_2$ and $L_3$ (Fig 1B).

## Clustering of 12 DLPFC samples with STGIC

The 12 DLPFC samples can be divided into 3 groups by shape and cluster number, each group containing 4 samples. The samples with sample id from 151507 to 151510 consist of 7 clusters and nearly all the spatial domains according to annotated labels look like straight strip, those from 151669 to 151672 consist of 5 clusters and all the spatial domains present gentle wave shape, those from 151673 to 151676 comprise 7 clusters and all the spatial domains present obvious arcs.

With a unified set of hyper-parameters during pre-processing, pre-clustering and clustering stages, the 12 DLPFC samples are clustered by STGIC with each slice once. Through clustering with DCF in STGIC, the mean and median ARIs are elevated to 0.58 and 0.60 from the level demonstrated by AGC pre-clustering. Although the general rising in performance is flawed by drops of ARI in several samples, it can be readily observed that the drops are so minuscule for most of the samples showing a lower ARI by clustering than by pre-clustering with only one striking exception of 151673 which demonstrates a significant ARI decrease (Fig 4A). For 151673, ARI has kept rising to 0.62 from 0.60 by the fourth iteration of DCF training, yet then goes through a continuous fall to 0.49 at the end, which may be imputed to the iteration termination mechanism adopted by STGIC, the function of DCF to extract feature, however, may still work. The trends of ARI curves for 151671 and 151675 unfold similar trends to 151673, but the drop of 0.01 is quite moderate. Only for 151508 and 151674, does STGIC not bring any ARI increases, but the drop remains within 0.03 (Fig 4A).

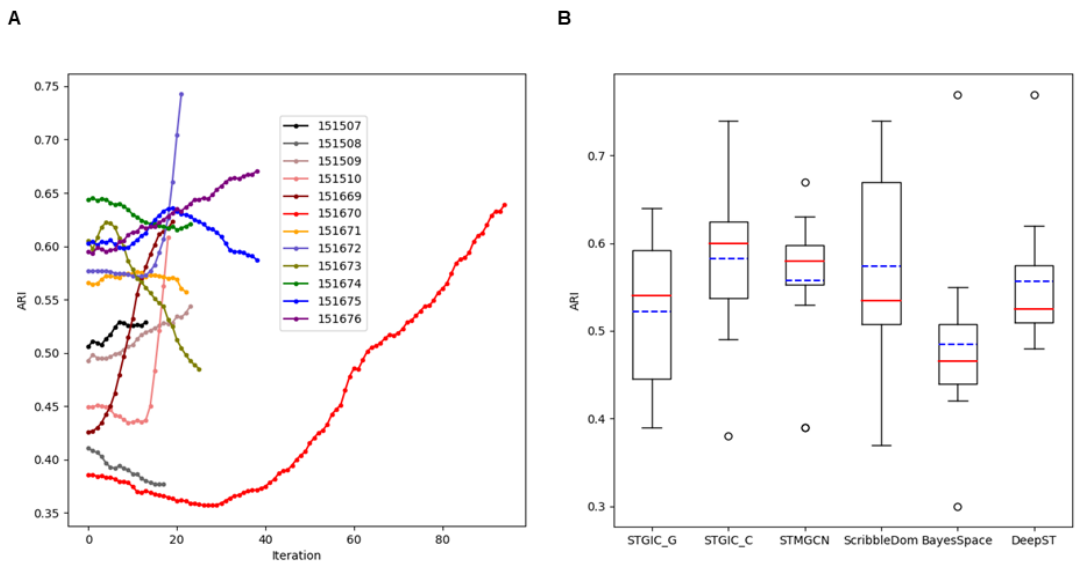

**Fig 4. Variation trend of ARI measuring STGIC performance and comparison**

**with baselines among the 12 DLPFCs.** A. ARI curves with iterations going on for every sample. At the $0^{th}$ iteration, the ARI is calculated with the pre-clustering result and then is with the cluster assignment generated in each iteration of DCF training. B. Boxplot showing ARIs computed with clustering labels generated by STGIC and baseline methods. "STGIC_G" represents the pre-training of AGC in STGIC and "STGIC_C" represents DCF which presents the ultimate clustering labels for STGIC. The blue dash lines and red solid lines respectively mark the positions of mean and median ARI.

AGC performs the best in the third group, but as far as this group is concerned, DCF tends to discount AGC's achievement except in 151676. By contrast, in the remaining two groups which AGC is not very adept at, DCF tends to significantly improve the pre-clustering performance. The phenomenon suggests AGC and DCF are individually suitable for different kinds of samples and the combination of them guarantees a high performance overall.

Our pre-clustering and clustering results are compared with four baseline methods, including STMGCN, ScribbleDom, BayesSpace and DeepST. The mean and median ARI of AGC pre-clustering in STGIC are respectively 0.52 and 0.54, suggesting the pre-clustering AGC performs better than only BayesSpace. However, the mean and median ARIs displayed by the final clustering of DCF in STGIC are the highest of all (Fig 4B), demonstrating the superiority in clustering accuracy of STGIC over all the baseline methods. The visualization of spatial domains in the 12 samples are depicted (Fig 5) according to the pre-clustering and clustering labels generated by STGIC.

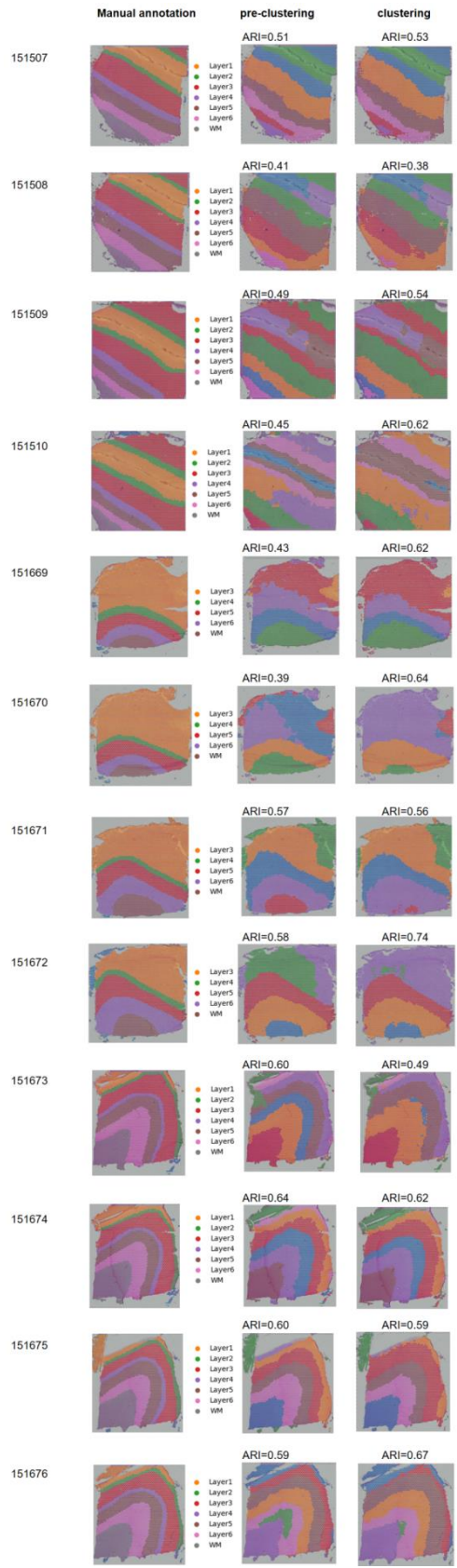

**Fig 5. Annotated spatial domains as well as those derived from pre-clustering**

**and clustering of STGIC.** The figure legends about the labels are only for the annotated spatial domains.

## Ablation study on DFC

To corroborate the necessity of adapting the training objectives adopted in the unsupervised image segmentation algorithm and TESLA, The KL divergence loss is removed and $q_{cut}$ is set to be 0 so that all spots are involved in calculating cross entropy. Besides, the closest neighboring spots pairs in diagonal direction are not considered any more for calculating of spatial continuity loss, correspondingly, this kind of training objective is restored completely to those adopted by TESLA and denoted as "ce + ortho". As a result, an awful performance is brought about with mean and median ARI demonstrated by DCF being 0.39 and 0.40. For individual samples, rising from pre-clustering ARI is only seen in 151507 and 151669 while the remaining samples suffer a slumping of ARI (Fig 6A). Restoration of spatial continuity loss in diagonal direction from the above situation doesn't make any obvious difference (Fig 6B), the kind of training objective is denoted as "ce + ortho + diag". If $q_{cut}$ is restored to 0.5, spots with high confidence pseudo-labels are screened for calculating cross entropy and spatial continuity loss in three directions are simultaneously calculated, the mode of training objective is denoted as "hce + ortho + diag", which raise the mean and median ARI to 0.45 and 0.41 (Fig 6C). The performance will be improved by a substantial extent if the self-supervision KL divergence loss is introduced to restore the integral training objectives of STGIC, as has been elucidated. However, to validate the positive effect of spatial continuity loss in the diagonal direction, the corresponding part is subtracted from the integral loss, and the mode of training objective is denoted as "KL + hce + ortho", which demonstrates a performance a little worse than the integral objective does, since the mean and median ARI are both 0.56 (Fig 6D). Although this mode leads the clustering ARI in some samples to be even higher than the integral objective does, it fails in getting an ideal performance in some other samples, especially in the sample 151670. To validate the reasonability of the adaptation of convolution kernel weights to spatial distance, the related operation is abolished, decreasing the mean and median ARI to 0.49, and the mode is denoted as "no considering distance". In the case, the clustering ARI is lower than the pre-clustering value in every sample except for 151669 and 151670 (Fig 6E). If the fixing of the weights at the 4 corners of the 3*3 kernel to be 0 is further canceled, the mean and median ARI decrease to 0.48 and 0.49, and the mode is denoted as "no considering distance + containing corners". Among the 12 samples, only the ARI of 151508 rises after clustering (Fig 6F).

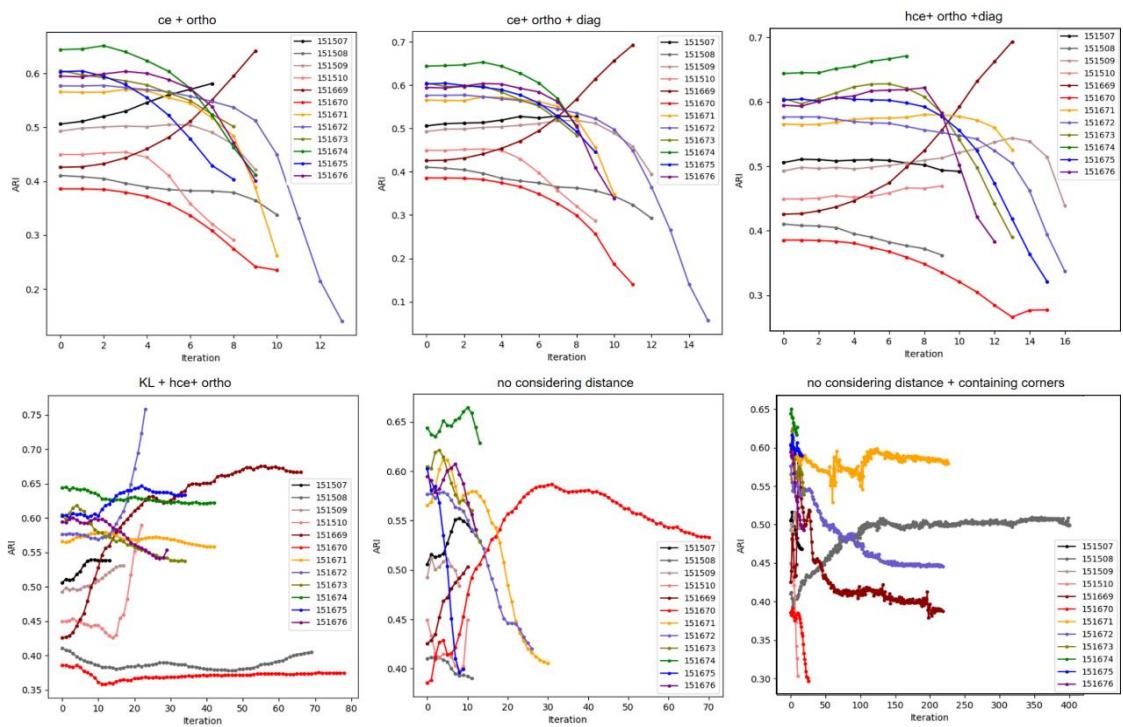

**Fig 6. Variation trend of ARIs in all cases of ablation study.**

Incidentally, we have also tried with several contrastive learning designed for graph data and intended to transfer these methods to our problem as we deal with the self-supervision KL divergence. These contrastive learning methods including GCA [27], DGI [28] and GIC [29], don't improve the performance significantly after integration into our existing training objective for DCF. Therefore, the framework of STGIC doesn't take any of these contrastive learning skills.

# STGIC can resolve finer spatial domains in 10x Visium human breast cancer

Manually annotated labels from SEDR's author has been adopted herein to measure our pre-clustering performance with STGIC [30]. According to the annotated labels, the sample is divided into 20 spatial domains, which consist of 5 ductal carcinomas in situ (DCIS) or lobular carcinomas in situ (LCIS) regions with prefix "DCIS/LCIS", 7 invasive ductal carcinoma (IDC) regions with prefix "IDC", 6 tumor edge regions with prefix "Tumor_edge" and 2 healthy regions with prefix "Healthy". Therefore, we pre-specify the cluster number to be 20 and use STGIC to cluster the sample. Our visualization of the resulting spatial domains demonstrates large overlaps with the main DCIS/LCIS and IDC regions, though some DCIS/LCIS and IDC regions are integrated into one domain in rare cases. Three small tumor edge regions with annotated labels "Tumor_edge_4", "Tumor_edge_5" and "Tumor_edge_6" have no corresponding spatial domains as well as the smaller healthy region "Healthy_2" as they are contained in domains corresponding to other regions (Fig 7A). In UMAP [31] visualization, neighboring domains in the above spatial domain visualization are also adjoining each other (Fig 7B). The similarity in domain distribution between the two types of visualization indicates the smoothing ability of STGIC to lead more similar embeddings among spatially closer spots. SVGs detection is carried out with the guidance of STGIC-identified spatial domains and the annotated spatial domains is used as positive control, besides SVGs are also detected with SPARK-X [32]. The intersection of SVG sets from STGIC-identified and the annotated spatial domains contains 149 genes with each set composed of 340 and 190 genes respectively. The set comprising the top 1000 SVGs identified by SPARK-X have only a few common elements with the former two sets. The commonly used metrics of Moran's I statistics for quantifying spatial autocorrelation of gene expression are calculated respectively within the three sets. SVG sets from the annotated and STGIC-identified labels get much higher Moran's I score than the set ascertained by SPARK-X (Fig 7C).

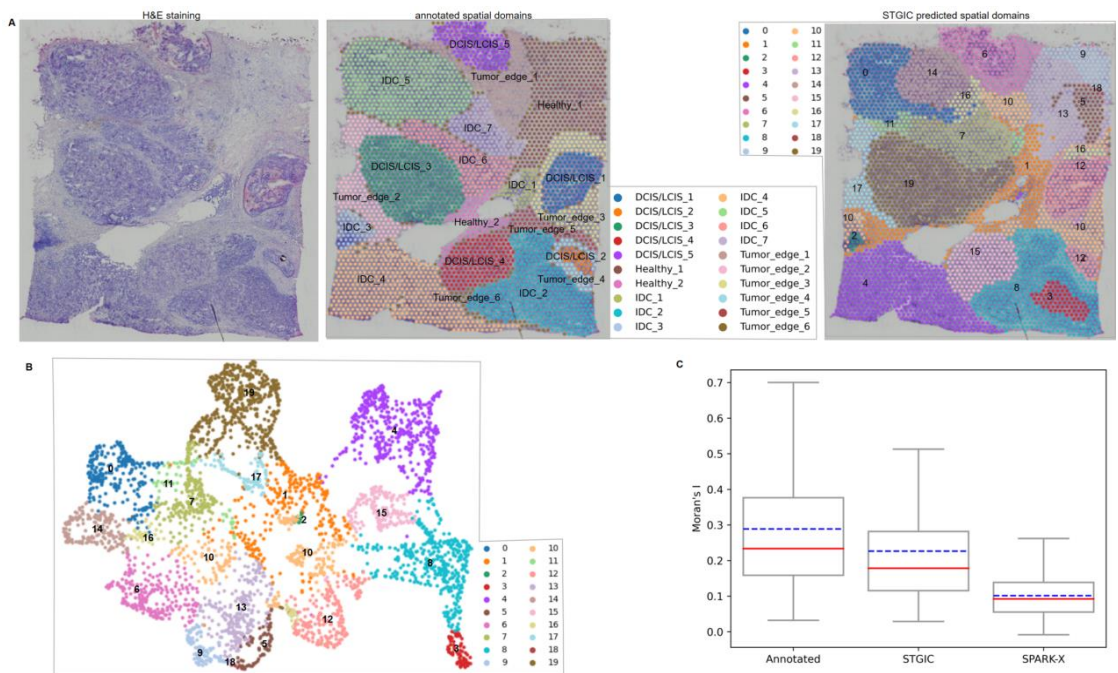

**Fig 7. Analysis of 10x Visium human breast cancer based on STGIC clustering.** A. H&E histological slice image, the manually annotated spatial domains and those identified by STGIC clustering. B. UMAP plot of spatial domains by STGIC clustering. C. Moran's I score of SVGs detected in the annotated and STGIC identified spatial domains as well as those detected by SPARK-X.

# STGIC can resolve finer spatial domains in 10x Visium mouse posterior brain

STGIC manages to depict the fine-grained spatial domain of the sagittal slice of mouse posterior brain with the pre-specified cluster number of 20, especially in cerebellar cortex. In reference to the corresponding anatomical structure presented by Allen Mouse Brain Atlas, cerebellar cortex can be divided into three layers from outer to inner which are molecular layer, granular layer and fiber tract. These layers correspond respectively to our spatial domains 18, 8 and 2 (Fig 8A). The neighborhood relationship between spatial domains is calculated and visualized as a neighborhood enrichment plot which highlights spatial approaching among the above three domains (Fig 8B). *Pcp2*, *Car8*, *Calb1* and *Ppp1r17* are marker genes abundantly expressed in Purkinje cells in cerebellar cortex [15, 33, 34]. *Pcp2* and *Car8* are detected as top SVGs of Domain 18, while *Calb1* and *Ppp1r17* are found to be top SVGs of Domain8. *Gabra6* [35] is marker genes highly expressed in granule neurons of cerebellum and detected as top SVGs of Domain 2. In addition to cerebellum, dentate gyrus in cerebrum is also depicted clearly by two of our spatial domains 13 and 19, in which *Hpca* and *C1ql2* [17, 36], markers of dentate gyrus, are identified as their top SVGs. The Cerebral cortex corresponds to two of our spatial domains 3 and 5, characterized respectively by SVGs *Gm11549* and *Hs3st2*, which are both marker genes of cerebral cortex [37, 38]. values of Moran's I are calculated with the SVGs sets derived from STGIC-identified spatial domains and with the sets composed of the top 1000 SVGs detected by SPARK-X, the mean and median of the former are much higher than those of the latter, the STGIC-derived sets have 1080 SVGs and has only 16 genes in common with SPARK-X's top 1000 SVGs set.

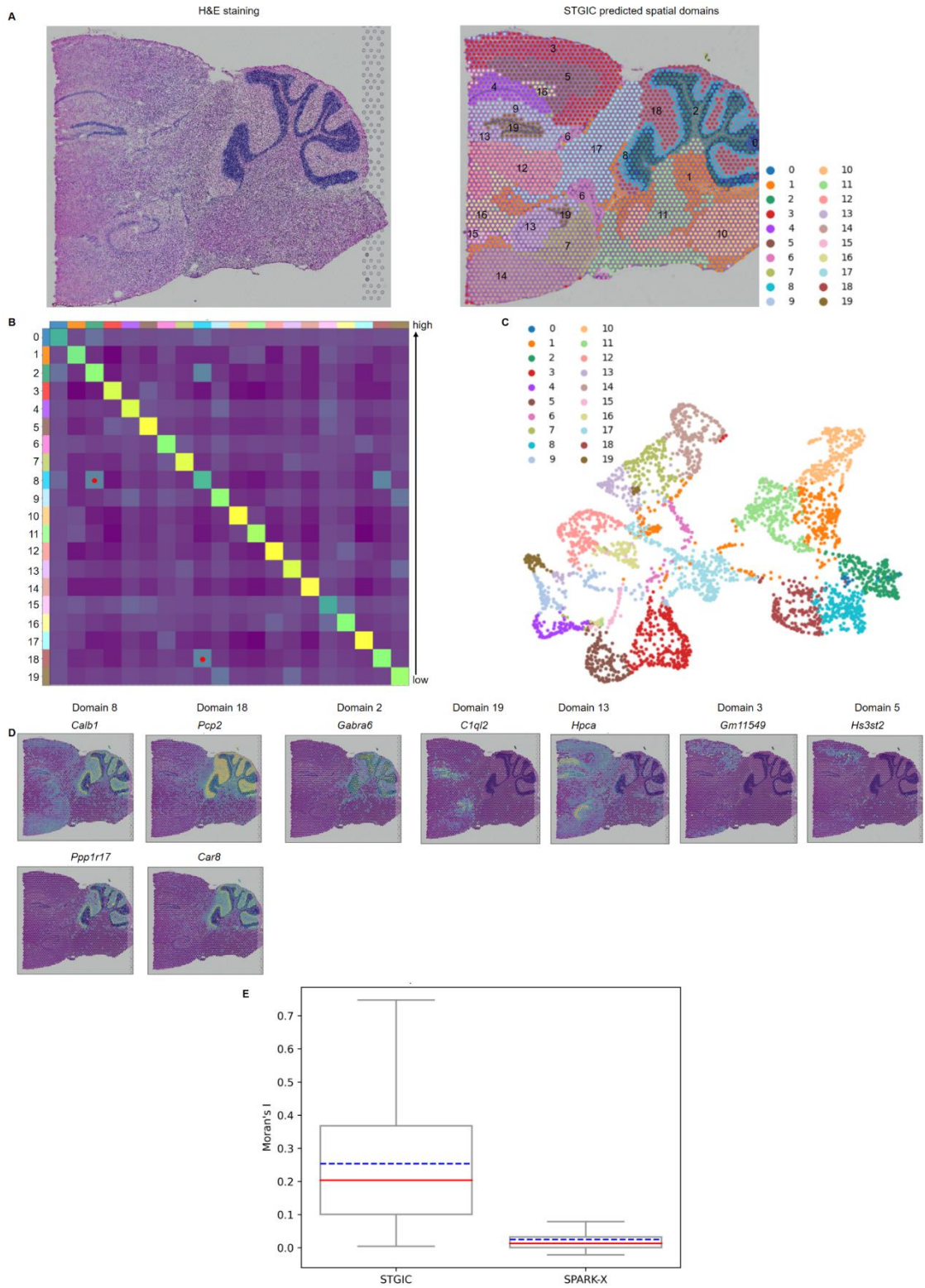

**Fig 8. Analysis of 10x Visium mouse post-brain sagittal slice based on STGIC clustering.** A. H&E histological slice image and visualization of Spatial domains identified by STGIC clustering. B. Neighborhood enrichment plot of spatial domains identified by STGIC. Two pairs of spatial domains having the strongest neighborhood

relationship are highlighted with red dot. C. UMAP plot of spatial domains by STGIC clustering. D. Spatial distribution of marker genes corresponding to layers in cerebellar cortex, cerebral dentate gyrus and cortex. Marker genes highly expressed in the same domain are arranged in the same column. E. Moran's I score of SVGs detected in STGIC identified spatial domains and those detected by SPARK-X.

The 10x Visium coronal slice of mouse posterior brain with manually annotated labels by Squidpy's author is also analyzed through STGIC [39]. The annotated spatial domains clearly delineate the anatomic structure of post-brain by delimiting regions of cortex, fiber tract, hippocampus, hypothalamus, thalamus, striatum and pyramidal layer including dentate gyrus. Following the annotation, cluster number is set to be 15 and the STGIC-identified clustering displays an ARI of 0.64. Our spatial domains correspond well to those of annotated ones, except for the integration of the outermost two layers of cortices into Domain 14 as well as the deficiency of Domain 7 and 11 in demarcating the thin pyramidal layer and dentate gyrus layer. However, the drawback of Domain 7 and 11 is moderate since they respectively cover the regions of the two layers so that it is not affected to detect their marker genes *Hpca* and *C1ql2* as top SVGs. Therefore, the two domains have their counterparts in Domain 13 and 9 in the sagittal data which feature the same marker genes. The UMP visualization separates our spatial domains widely apart but still keeps neighboring spatial domains closer than those not adjoining each other in spatial domain visualization. Scores of Moran's I are calculated as to three SVGs sets derived respectively from annotated spatial domains, STGIC-identified domains and SPARK-X. The former two sets contain 1169 and 695 genes and have 571genes in common, both of them have a negligible quantity of genes contained in the top 1000 SVGs detected by SPARK-X. The former two sets get their scores of Moran's I much higher than the last.

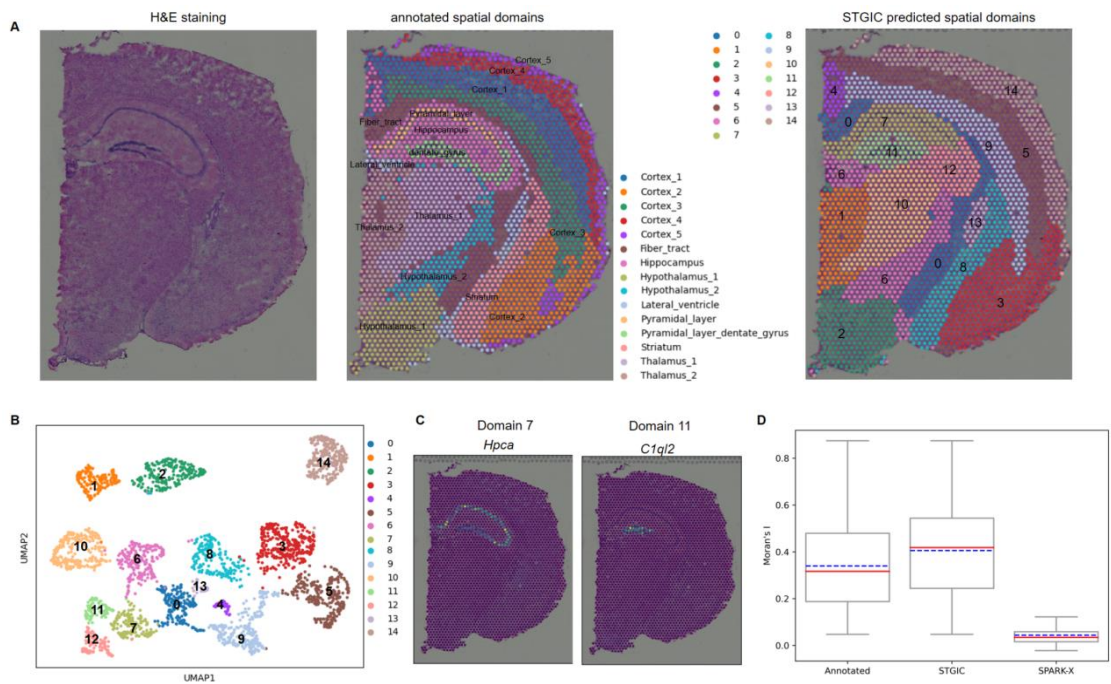

**Fig 9. Analysis of 10x Visium mouse post-brain coronal slice based on STGIC clustering.** A. H&E histological slice image, the manually annotated spatial domains and those identified by STGIC clustering. B. UMAP plot of spatial domains by STGIC clustering. C. Spatial distribution of marker genes corresponding to layers in cerebral dentate gyrus. D. Moran's I score of SVGs detected in the annotated and STGIC identified spatial domains as well as those detected by SPARK-X.

# Clustering derived from STGIC explicitly delineate the laminate structure of Stereo-seq mouse olfactory bulb

The Stereo-seq data is of high resolution at cellular level and have 19109 spots. Mouse olfactory bulb mainly consists of 7 layers which from the inner to the outer are RMS, GCL, IPL, MCL, EPL, GL and ONL. These layers have been labeled in Allen Brain Atlas [40] and also by SEDR's author [30]. Every layer has at least one marker gene. When the cluster number is set to be 9, each layer has its own corresponding spatial domain identified by STGIC. Domain 5, 2, 1, 7, 0, 6, 4 are mapped to the above 7 layers in turn and marker genes of these layers have also been found out in top SVGs of the corresponding spatial domains, which proves the precision of STGIC clustering indirectly. To further validate the capability of STGIC, the resolution is decreased by binning the data to reserve 14781 spots. STGIC still manages to demarcates a majority of the anatomical layers except that EPL and GL are merged into Domain 1, as is demonstrated by cooccurrence of the marker genes *Slc6a11* and *Cck* [17] of the two layers in the SVGs of this domain. Since cells abundantly expressing the two genes are so close in spatial distribution as to be easily binned together, it is expected that a unified single domain displays high expression of the two genes.

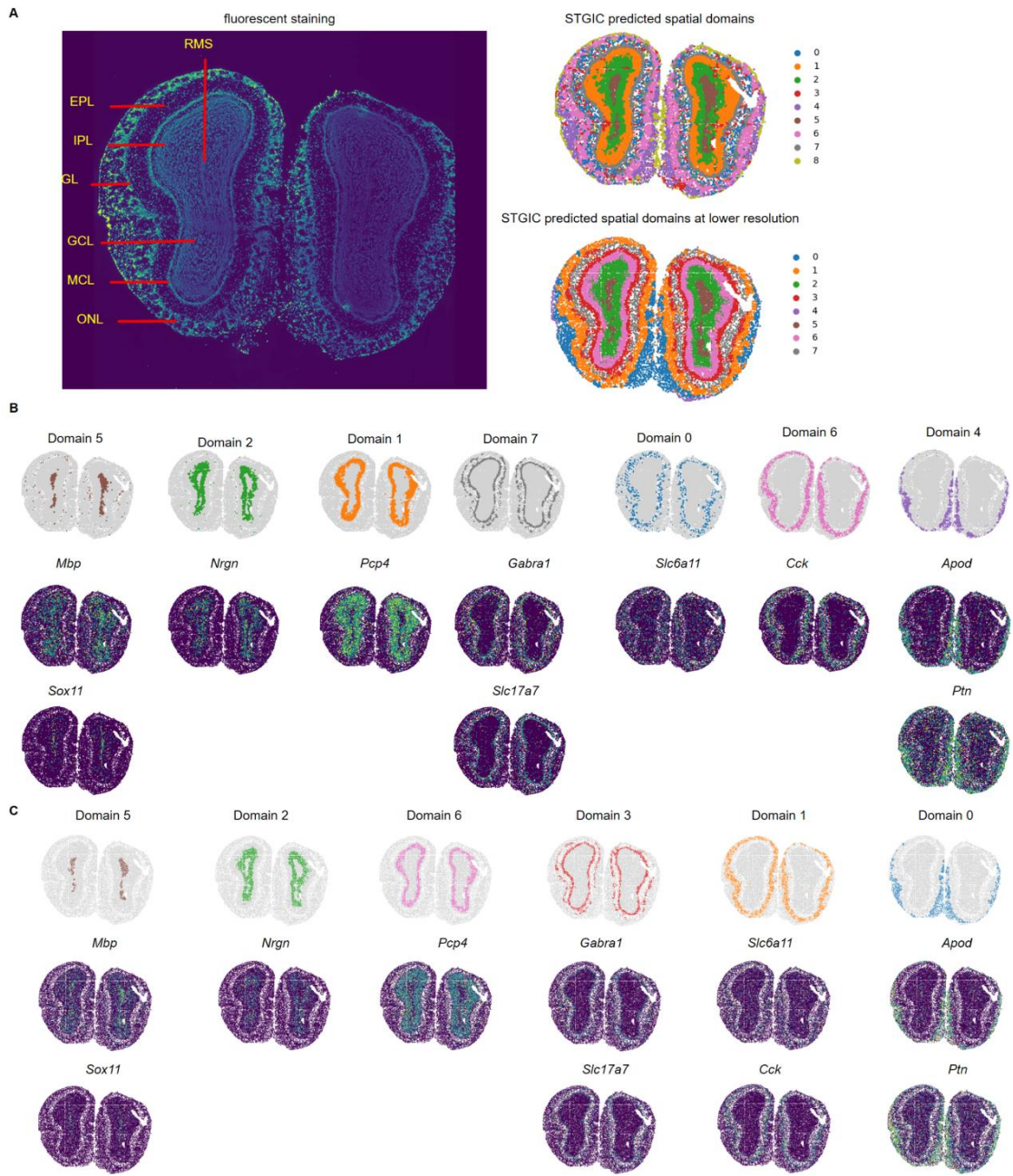

**Fig 10. STGIC depicts layers of Stereo-seq mouse olfactory bulb with great clarity.** A. Manually annotated layers of mouse olfactory bulb and visualization of spatial domains identified by STGIC on data with cell-level and lower resolution. B. Visualization of individual domains and spatial distribution of their marker genes for the cluster assignment as to data with cell-level resolution by STGIC. C. Visualization of individual domains and the spatial distribution of their marker genes for the cluster assignment as to data with lower resolution by STGIC. Marker genes are arranged in the same column as the corresponding domains in both B and C.

## Conclusions

We have developed a deep learning method STGIC for ST clustering. STGIC combines graph convolution AGC and an image convolution framework DCF together to attain high clustering performance on multiple datasets. It starts with AGC pre-clustering to obtain high quality pseudo-labels and takes advantage of special setup of convolution kernels in DCF to extract features from a virtual image derived from regular lattice on the chip of 10x Visium and Stereo-seq, making sure that all neighbor spots only in a certain distance are involved in updating the feature of each spot and contributions of these neighbors to the updated feature are correlated with spatial distance. The training of DCF doesn't simply follow any existing method in ST clustering, but transfers the self-supervision skill originally for graph learning to our image field, the spatial continuity loss and pseudo-labels skill adopted by the forgoing unsupervised image segmentation algorithm are modified. The spatial continuity loss considers diagonal direction and the pseudo-labels skill is limited only to spots with high confidence pseudo-labels.

STGIC demonstrates SOTA precision on the benchmark of the DLPFC dataset consisting of 12 samples. Its value is also displayed by the capability of depicting fine structure of human breast cancer, mouse post-brain and olfactory bulb, the ability to guide the identification of marker genes through the predicted spatial domains as well as the expandability from 10x Visium to Slide-seq.

# Method

## Overview of datasets

All data analyzed in this paper are available in raw form from their original authors.

(1) The 10x Visium Human DLPFC dataset is available within the spatialLIBD package (http://spatial.libd.org/spatialLIBD). (2) The 10x Visium Human breast cancer dataset is downloaded from the website (https://support.10xgenomics.com/spatial-gene-expression/datasets). The manually annotated labels is recorded at website https://github.com/JinmiaoChenLab/SEDR_analyses/blob/master/data/BRCA1/metadata.tsv. (3) The 10x Visium MouseBrain dataset of sagittal slice is dowloaded from the website https://support.10xgenomics.com/spatial-gene-expression/datasets. (4) The 10x Visium MouseBrain dataset of coronal slice with manually annotated label is downloaded by operating the function squidpy.datasets.visium_hne_adata in the Python package squidpy. (5) The processed Stereo-seq mouse olfactory bulb tissue with cell-level resolution is downloaded from the website https://github.com/JinmiaoChenLab/SEDR_analyses. The data with lower resolution is binned by ourselves.

## Formal details of STGIC

    **A. Computation of Adjacency *A* for AGC input.** Following the practice of SpaGCN to generate the self-looped adjacency from distance matrix, we derive the adjacency in the formula:

$$A_{ij} = e^{-\frac{d_{ij}^2}{2\sigma^2}} \qquad (1)$$

$A_{ij}$ is the element of the adjacency $A$ in row $i$ and column $j$, $d_{ij}$ is the element of the distance matrix in row $i$ and column $j$ which represents the Euclidean distance between the $i^{th}$ and $j^{th}$ spot, σ is a parameter identified by searching a series of values to make the mean of row sums of the difference between $A$ and identify matrix equal a pre-specified hyper-parameter which is set to be 0.5 herein by default [15].

**B. Conversion of transcription and spatial information to virtual image $X$.** For 10x Visium data, $X$ is generated by arranging each spot at the lattice coordinates rather than pixel coordinates since the spatial distribution would be too sparse for neighboring spots to be captured in one receptive field simultaneously if the pixel coordinates were adopted. The maximum abscissa and ordinate respectively adding 1 are the height and width of $X$ as the coordinates start with 0 rather than 1. PCA is carried out as to the preprocessed gene transcription matrix $M$ in the range of the top 3000 highly variable genes and the top 15 PCs are taken as pixel values of all spot pixels. The resulting 3000 dimensions of spots-averaged vector of the matrix $M$ and the top 15 eigen-vectors of the above PCA are used to compute pixel values of the null and background pixels by subtracting the spots-averaged vector from a 3000-dimension zero vector and then computing the top 15 PCs with the mentioned top15 eigen-vectors. The imputation for background pixel is reasonable in that no cells exist in the background area and thus no expression of any genes, namely the gene expression is represented with zero vector. Our practice is to reduce zero vector to 15 dimensional with the eigen-vectors and average vector, which however seems inappropriate to process null pixels since there could have been cells in these area except for technical limitation on spatial resolution to cause the omission and thus gene expression in these areas should not be taken as 0. Fortunately, this seemingly implausible point is circumvented by our introducing of dilated convolution kernel in DCF which ensuring the feature extraction of each spot pixel only refers to the neighboring spot pixels, eliminating the impact of the neighboring null pixels. Besides, only spot pixels are extracted from DCF-generated feature image to compute the value of loss function. Exclusion of background pixels and null pixels from loss calculation further attenuates the impact of the two kinds of unwanted pixels. For Stereo-seq data, only pixel coordinates are available, and hence *X* is constructed with pixel coordinates. Unlike the case in 10x Visium, the process will not cause sparse spatial distribution among spot pixels for the high resolution. Besides, closest spot pixels are not separated by null pixels because of the lattice structure. Therefore, we need only to compute pixel values for spot pixels and null pixels with the same method described above.

### C. Obtaining of pre-clustering label $C_0$ by AGC.

The smoothed embedding matrix $H_t \in R^{s \times 50}$ after iterations of *t* times from the initial feature matrix *F* is expressed as:

$$H_t = (I - L/\lambda)^t F \qquad (2)$$

where *I* is identity matrix, $\lambda$ is the approximate to the largest eigen-value of the symmetrically normalized Laplacian matrix *L*. we follow the practice of AGE to set $\lambda$ to be 1.5 [41].

In each iteration, the smoothed embedding matrix multiplying its own transpose generates a similarity matrix, the eigen-vector matrix of which is used to carry out spectral clustering. Specifically, the top 26 eigen-vectors provide each of the *s* spots with a 26-dimension representation vector, based on which Clustering is performed with Kmeans to generate a cluster label assignment. The assignment $C_t \in Z^{1 \times s}$ generated at the $t^{th}$ iteration is used to calculate the intra-cluster distance *Intra*($C_t$) as:

$$Intra(C_t) = \frac{1}{|C_t|}\sum_{l \in C_t} \frac{1}{|l|(|l|-1)} \sum_{\substack{i,j \in l \\ i \neq j}} ||H_t^i - H_t^j||_2 \tag{3}$$

where $|C_t|$ is the number of different cluster labels in $C_t$, $l$ is any of different labels in $C_t$, $|l|$ is the number of spots assigned to the label $l$, $i$ and $j$ are any pair of spots with different indices and assigned to the label $l$, $H^i_t$ and $H^j_t$ are respectively the smoothed embeddings of the $i^{th}$ and $j^{th}$ *spots* in the $t^{th}$ iteration, $||\ ||_2$ is L2-norm. Once the intra-cluster distance begins to increase, iteration will be terminated and cluster assignment obtained in the iteration immediately before the increase are taken as pre-clustering label $C_0$.

**D. Construction of DCF for feature extraction.** Whether 10x Visium or Stereo-seq data is to be analyzed, DCF functions by its two sub-frameworks. As has been elucidated, the sub-frameworks contain two different kinds of convolution kernels. For 10x Visium data, the convolution kernels both have dilation rate of 2, but have different kernel sizes respectively of 3*3 and 2*2. The kernel weights at the four corners of the 3*3 convolution kernels are kept to be zero constantly to mask spots corresponding to the corners of receptive fields, since the spots in a receptive field acted upon by them are too far from the center to play an appreciable role in updating the feature of the center spot relatively to other spots in the same receptive field (Fig 2B). Besides, after updating the kernel weights through back-propagation during each iteration, adaptation of the kernel weights is implemented to the distance from the position of kernel elements to that of kernel centers such that same weight values are shared among elements equally far from the centers (Fig 2B). The ultimate feature image is obtained by summing up the output feature images from the two sub-frameworks by certain weights which are set to be 0.7 for the one containing 3*3 kernels and 0.3 for the other. For Stereo-seq, it is noteworthy that spots are arranged continuously on the lattice so that spot pixels are distributed without any separation by null pixels in the virtual image *X* (Fig 11A). Correspondingly, we adjust the adaptation method of kernel weights to the characteristic of Stereo-seq. Convolution kernels with kernel size 3*3 and dilation rate respectively of 1 (no dilation at all) and 2 are respectively adopted in the two sub-frameworks. In addition to weights at the four corners of the dilated kernel with size 3*3 being kept zero all the way, the weight at the center should also be fixed at zero, since the center of receptive fields is paid attention to by the kernel without dilation and repetitive consideration of the center would otherwise overestimate its importance for updating the information of itself (Fig 11B). When the outputs from the two sub-frameworks are sum up to get the ultimate feature image, the pre-specified weight for the one from the sub-framework depending on kernels without dilation are set to be 0.9 and the other 0.1. For 10x Visum and Stereo-seq data, the feature images output by the first three blocks of all the sub-frameworks are set to have 100 channels and the feature image output by the last blocks of the sub-frameworks are set to have the channel number equal to the cluster number *n*.

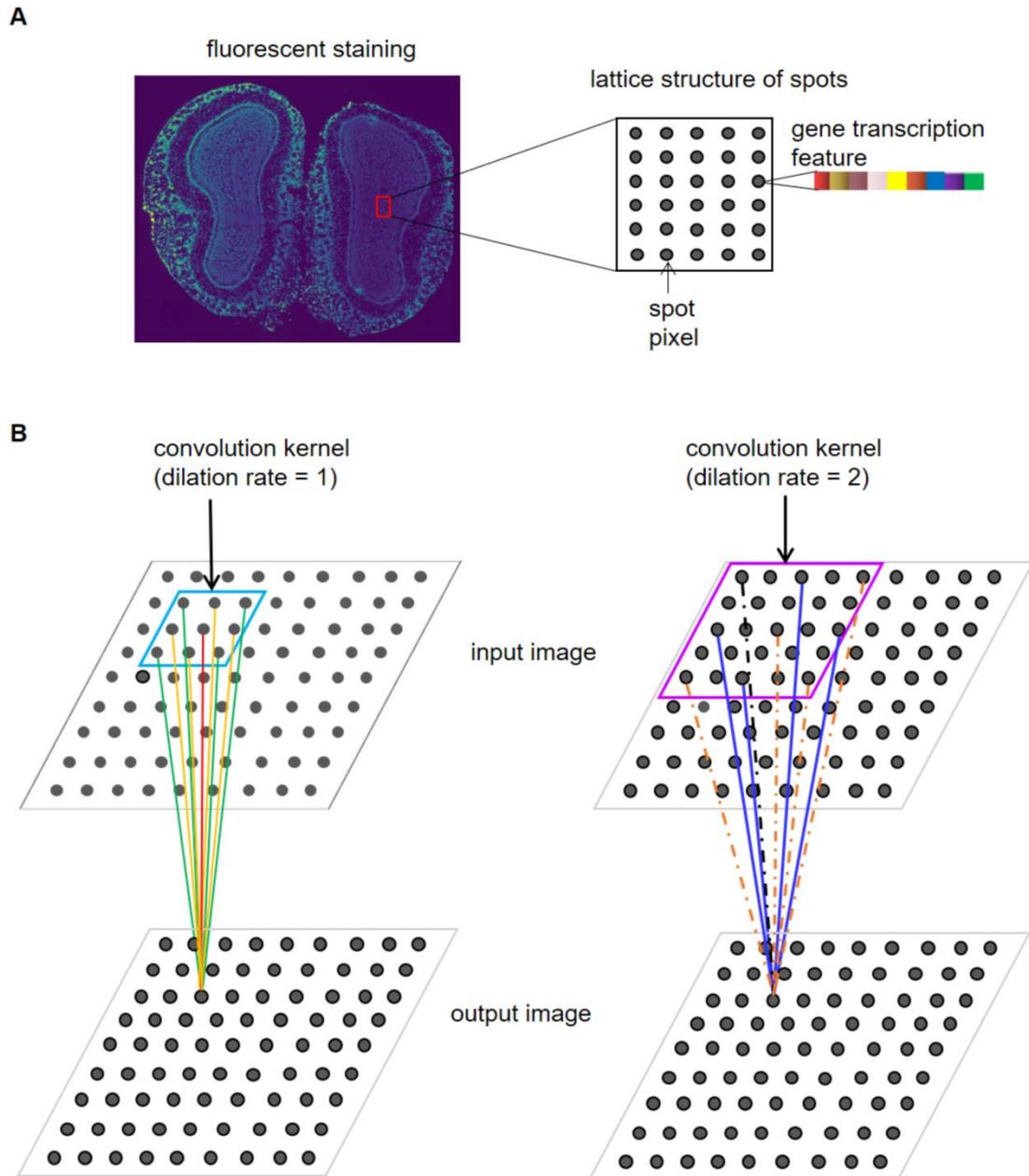

**Fig 11. Lattice structure of Stereo-seq and the adaptation of convolution kernel to the structure.** A. Spots are arranged one by one regularly in both longitudinal and transverse directions, quite different from what is adopted by 10x Visium on an alternate basis. B. Two kinds of convolution kernels are used in DCF, both have kernel size of 3*3, but one has dilation rate of 1 and the other has that of 2. Convolution kernel shares the same weight at positions with equal distance to the center. Lines with the same color represent the same kernel weight when extracting feature from a receptive field. Orange dash lines represent the weight of zero to ignore these spots during feature extraction.

**E. Self-supervision training of DCF.** A self-supervision trick realized via KL divergence firstly proposed by the community detection algorithm SDCN [42] and then employed by SpaGCN [15] is also used in our DCF. Given that the trick is as such designed for graph node embedding generated by GCN, we have to calculate with the *E* matrix composed of embedding of all spots extracted from the feature image of *O* instead of directly using it. The *Q* distribution is calculated as:

$$q_{ij} = \frac{(1+||E_i-\mu_j||_2^2)^{-1}}{\sum_{j'=1}^{|C|}(1+||E_i-\mu_{j'}||_2^2)^{-1}} \qquad (4)$$

where $q_{ij}$ is the probability of the $i^{th}$ spot belonging to the $j^{th}$ cluster, $E_i$ is the embedding of the $i^{th}$ spot in $i^{th}$ row of the matrix *E*, $\mu_j$ is the representation vector of the $j^{th}$ cluster centroid in the $j^{th}$ row of the matrix $\mu$, $|C|$ is the number of different labels in the cluster assignment matrix *C*, $||\ ||_2$ is L2-norm. Based on *Q*, the distribution *P* is calculated as:

$$p_{ij} = \frac{q_{ij}^2/\sum_{i'} q_{i'j}}{\sum_{j'}(q_{ij'}^2/\sum_{i'} q_{i'j'})} \qquad (5)$$

where the subscripts have the same meaning with those in formula (4).

**F. Learning rates for pre-training and training of DCF.** Learning rates during DCF pre-training and training are 0.05 and 0.01.

**G. Termination condition for pre-training and training of DCF.** The maximum iteration number of pre-training and training are 200 and 400. Besides, the pre-training of DCF will stop if either of the following two cases occur: (*i*) at least one cluster exists in current iteration having the member spots account for less than a certain fraction (set to be 0.01 for the 12 DLPFC benchmark) of the total, (*ii*) current iteration experience the updating of the matrix *P* and the ratio of spots having different cluster labels from those in last iteration to the total spots number $s$ is lower than 0.001.

**H. Weights of all parts of the loss for training DCF.** The spatial continuity loss $L_3$ is the weighted sum of two parts, the first is computed with the closest spots in the orthogonal (including longitudinal and transverse) direction and the second in the diagonal direction. Their weights are respectively 0.62 and 0.58. The total loss $L_t$ is the weighted sum of $L_1$, $L_2$ and $L_3$ with their weights of 0.78, 0.71 and 1.

## Identifying spatially variable genes

For STGIC-identified and annotated spatial domains, the python package SCANPY [43] is used to detect SVGs in each domain with the function scanpy.tl.rank_genes_groups implementing Wilcoxon test. Simultaneously, SVGs are also identified with the algorithm SPARK-X implemented by the R package SPARK.

## Neighborhood enrichment of spatial domains

The function of squidpy.gr.nhood_enrichment from the python package SQUIDPY [39] is used to calculate the extent to which any two spatial domains are neighboring to each other.

## Calculation of spatial autocorrelation of gene expression

The autocorrelation is measured by Moran's I score and implemented with the function squidpy.gr.spatial_autocorr from the package SQUIDPY.

# Calculation of clustering performance

When annotated labels is provided, clustering performance can be computed with the annotated and predicted labels. Suppose $U = \{U_1, U_2, ..., U_{cn}\}$ are the set of annotated labels containing $c_n$ different elements and $V = \{V_1, V_2, ..., V_{cp}\}$ are the set of predicted labels containing $c_p$ different elements, clustering performance is measured by ARI, which is calculated as:

$$ARI = \frac{\Sigma_{i,j}\binom{n_{ij}}{2} - \frac{\Sigma_i\binom{n_{i.}}{2}\Sigma_j\binom{n_{.j}}{2}}{\binom{n}{2}}}{\frac{1}{2}[\Sigma_i\binom{n_{i.}}{2}+\Sigma_j\binom{n_{.j}}{2}] - \frac{\Sigma_i\binom{n_{i.}}{2}\Sigma_j\binom{n_{.j}}{2}}{\binom{n}{2}}} \quad (6)$$

where $n_{i.}$ and $n_{.j}$ are the number of spots categorized as $U_i$ and $V_j$. $n_{ij}$ is the number of spots simultaneously attached to $U_i$ and $V_j$. The higher ARI score is, the more consistent the predicted labels are with the annotated ones.

PMID: 30914743; PubMed Central PMCID: PMCPMC6435756 community detection algorithms in commercial products and services.

4. Blondel VD, Guillaume J-L, Lambiotte R, Lefebvre E. Fast unfolding of communities in large networks. Journal of Statistical Mechanics: Theory and Experiment. 2008;2008(10). doi: 10.1088/1742-5468/2008/10/p10008.

5. Li X, Wang CY. From bulk, single-cell to spatial RNA sequencing. Int J Oral Sci. 2021;13(1):36. Epub 2021/11/17. doi: 10.1038/s41368-021-00146-0. PubMed PMID: 34782601; PubMed Central PMCID: PMCPMC8593179.

6. Marx V. Method of the Year 2020: spatially resolved transcriptomics. Nat Methods. 2021;18(1):1. Epub 2021/01/08. doi: 10.1038/s41592-020-01042-x. PubMed PMID: 33408396.

7. Ji AL, Rubin AJ, Thrane K, Jiang S, Reynolds DL, Meyers RM, et al. Multimodal Analysis of Composition and Spatial Architecture in Human Squamous Cell Carcinoma. Cell. 2020;182(6):1661-2. Epub 2020/09/19. doi: 10.1016/j.cell.2020.08.043. PubMed PMID: 32946785; PubMed Central PMCID: PMCPMC7505493.

8. Stickels RR, Murray E, Kumar P, Li J, Marshall JL, Di Bella DJ, et al. Highly sensitive spatial transcriptomics at near-cellular resolution with Slide-seqV2. Nat Biotechnol. 2021;39(3):313-9. Epub 2020/12/09. doi: 10.1038/s41587-020-0739-1. PubMed PMID: 33288904; PubMed Central PMCID: PMCPMC8606189.

9. Rodriques SG, Stickels RR, Goeva A, Martin CA, Murray E, Vanderburg CR, et al. Slide-seq: A scalable technology for measuring genome-wide expression at high spatial resolution. Science. 2019;363:1463-7. doi: 10.1126/science.aaw1219.

10. Wei X, Fu S, Li H, Liu Y, Wang S, Feng W, et al. Single-cell Stereo-seq reveals induced progenitor
37